\documentclass[11pt]{article}
\usepackage{epsfig}
\usepackage{amsmath}
\usepackage{hhline}
\usepackage{amssymb}
\usepackage{cite}

\newlength{\dinwidth}
\newlength{\dinmargin}
\begin{document}  
\def\mbig#1{\mbox{\rule[-2. mm]{0 mm}{6 mm}#1}}
%
%
\newcommand{\s}{\mbox{$s$}}
\newcommand{\ttra}{\mbox{$t$}}
\newcommand{\modt}{\mbox{$|t|$}}
\newcommand{\eminpz}{\mbox{$E-p_z$}}
\newcommand{\eminpzs}{\mbox{$\Sigma(E-p_z)$}}
\newcommand{\rap}{\ensuremath{\eta^*} }
\newcommand{\W}{\mbox{$W$}}
\newcommand{\w}{\mbox{$W$}}
\newcommand{\Q}{\mbox{$Q$}}
\newcommand{\q}{\mbox{$Q$}}
\newcommand{\xB}{\mbox{$x$}}  
\newcommand{\xF}{\mbox{$x_F$}}  
\newcommand{\xg}{\mbox{$x_g$}}  
\newcommand{\xbj}{x}
\newcommand{\xpom}{x_{\PO}}
\newcommand{\y}{\mbox{$y~$}}
\newcommand{\Qsq}{\mbox{$Q^2$}}
\newcommand{\qsq}{\mbox{$Q^2$}}
\newcommand{\kjet}{\mbox{$k_{T\rm{jet}}$}}
\newcommand{\xjet}{\mbox{$x_{\rm{jet}}$}}
\newcommand{\Ejet}{\mbox{$E_{\rm{jet}}$}}
\newcommand{\thjet}{\mbox{$\theta_{\rm{jet}}$}}
\newcommand{\pjet}{\mbox{$p_{T\rm{jet}}$}}
\newcommand{\et}{\mbox{$E_T~$}}
\newcommand{\kt}{\mbox{$k_T~$}}
\newcommand{\ptrans}{\mbox{$p_T~$}}
\newcommand{\pth}{\mbox{$p_T^h~$}}
\newcommand{\pte}{\mbox{$p_T^e~$}}
\newcommand{\ptsq}{\mbox{$p_T^{\star 2}~$}}
\newcommand{\as}{\mbox{$\alpha_s~$}}
\newcommand{\ycut}{\mbox{$y_{\rm cut}~$}}
\newcommand{\gx}{\mbox{$g(x_g,Q^2)$~}}
\newcommand{\xpart}{\mbox{$x_{\rm part~}$}}
\newcommand{\mrsdm}{\mbox{${\rm MRSD}^-~$}}
\newcommand{\mrsdmp}{\mbox{${\rm MRSD}^{-'}~$}}
\newcommand{\mrsdn}{\mbox{${\rm MRSD}^0~$}}
\newcommand{\lambdams}{\mbox{$\Lambda_{\rm \bar{MS}}~$}}
%
%
\newcommand{\gp}{\ensuremath{\gamma}p }
\newcommand{\gammasp}{\ensuremath{\gamma}*p }
\newcommand{\gammap}{\ensuremath{\gamma}p }
\newcommand{\gsp}{\ensuremath{\gamma^*}p }
\newcommand{\dsiget}{\ensuremath{{\rm d}\sigma_{ep}/{\rm d}E_t^*} }
\newcommand{\dsigrap}{\ensuremath{{\rm d}\sigma_{ep}/{\rm d}\eta^*} }
\newcommand{\epem}{\mbox{$e^+e^-$}}
\newcommand{\ep}{\mbox{$ep~$}}
\newcommand{\epl}{\mbox{$e^{+}$}}
\newcommand{\emi}{\mbox{$e^{-}$}}
\newcommand{\epm}{\mbox{$e^{\pm}$}}
\newcommand{\se}{section efficace}
\newcommand{\ses}{sections efficaces}
%
%
\newcommand{\phib}{\mbox{$\varphi$}}
\newcommand{\rh}{\mbox{$\rho$}}
\newcommand{\rhz}{\mbox{$\rh^0$}}
\newcommand{\ph}{\mbox{$\phi$}}
\newcommand{\om}{\mbox{$\omega$}}
\newcommand{\jpsi}{\mbox{$J/\psi$}}
\newcommand{\pipi}{\mbox{$\pi^+\pi^-$}}
\newcommand{\pip}{\mbox{$\pi^+$}}
\newcommand{\pim}{\mbox{$\pi^-$}}
\newcommand{\kk}{\mbox{K^+K^-$}}
\newcommand{\bsl}{\mbox{$b$}}
\newcommand{\alp}{\mbox{$\alpha^\prime$}}
\newcommand{\alpom}{\mbox{$\alpha_{\PO}$}}
\newcommand{\alpomp}{\mbox{$\alpha_{\PO}^\prime$}}
\newcommand{\rzzzz}{\mbox{$r_{00}^{04}$}}
\newcommand{\rzqzz}{\mbox{$r_{00}^{04}$}}
\newcommand{\rzquz}{\mbox{$r_{10}^{04}$}}
\newcommand{\rzqumu}{\mbox{$r_{1-1}^{04}$}}
\newcommand{\ruuu}{\mbox{$r_{11}^{1}$}}
\newcommand{\ruzz}{\mbox{$r_{00}^{1}$}}
\newcommand{\ruuz}{\mbox{$r_{10}^{1}$}}
\newcommand{\ruumu}{\mbox{$r_{1-1}^{1}$}}
\newcommand{\rduz}{\mbox{$r_{10}^{2}$}}
\newcommand{\rdumu}{\mbox{$r_{1-1}^{2}$}}
\newcommand{\rcuu}{\mbox{$r_{11}^{5}$}}
\newcommand{\rczz}{\mbox{$r_{00}^{5}$}}
\newcommand{\rcuz}{\mbox{$r_{10}^{5}$}}
\newcommand{\rcumu}{\mbox{$r_{1-1}^{5}$}}
\newcommand{\rsuz}{\mbox{$r_{10}^{6}$}}
\newcommand{\rsumu}{\mbox{$r_{1-1}^{6}$}}
\newcommand{\rzqik}{\mbox{$r_{ik}^{04}$}}
\newcommand{\rhzik}{\mbox{$\rh_{ik}^{0}$}}
\newcommand{\rhqik}{\mbox{$\rh_{ik}^{4}$}}
\newcommand{\rhaik}{\mbox{$\rh_{ik}^{\alpha}$}}
\newcommand{\rhzzz}{\mbox{$\rh_{00}^{0}$}}
\newcommand{\rhqzz}{\mbox{$\rh_{00}^{4}$}}
\newcommand{\raik}{\mbox{$r_{ik}^{\alpha}$}}
\newcommand{\razz}{\mbox{$r_{00}^{\alpha}$}}
\newcommand{\rauz}{\mbox{$r_{10}^{\alpha}$}}
\newcommand{\raumu}{\mbox{$r_{1-1}^{\alpha}$}}

\newcommand{\R}{\mbox{$R$}}
\newcommand{\rzero}{\mbox{$r_{00}^{04}$}}
\newcommand{\rone}{\mbox{$r_{1-1}^{1}$}}
\newcommand{\costh}{\mbox{$\cos\theta$}}
\newcommand{\cosp}{\mbox{$\cos\psi$}}
\newcommand{\costop}{\mbox{$\cos(2\psi)$}}
\newcommand{\cosd}{\mbox{$\cos\delta$}}
\newcommand{\cossqp}{\mbox{$\cos^2\psi$}}
\newcommand{\cossqt}{\mbox{$\cos^2\theta^*$}}
\newcommand{\sint}{\mbox{$\sin\theta^*$}}
\newcommand{\sintot}{\mbox{$\sin(2\theta^*)$}}
\newcommand{\sinsqt}{\mbox{$\sin^2\theta^*$}}
\newcommand{\costhst}{\mbox{$\cos\theta^*$}}
\newcommand{\vep}{\mbox{$V p$}}
\newcommand{\mpipi}{\mbox{$m_{\pi^+\pi^-}$}}
\newcommand{\mkk}{\mbox{$m_{KK}$}}
\newcommand{\mkaka}{\mbox{$m_{K^+K^-}$}}
\newcommand{\mpp}{\mbox{$m_{\pi\pi}$}}       
\newcommand{\mppsq}{\mbox{$m_{\pi\pi}^2$}}   
\newcommand{\mpi}{\mbox{$m_{\pi}$}}          
\newcommand{\mrho}{\mbox{$m_{\rho}$}}        
\newcommand{\mrhosq}{\mbox{$m_{\rho}^2$}}    
\newcommand{\Gmpp}{\mbox{$\Gamma (\mpp)$}}   
\newcommand{\Gmppsq}{\mbox{$\Gamma^2(\mpp)$}}
\newcommand{\Grho}{\mbox{$\Gamma_{\rho}$}}   
\newcommand{\grho}{\mbox{$\Gamma_{\rho}$}}   
\newcommand{\Grhosq}{\mbox{$\Gamma_{\rho}^2$}}   
%
%
\newcommand{\cm}{\mbox{\rm cm}}
\newcommand{\GeV}{\mbox{\rm GeV}}
\newcommand{\gev}{\mbox{\rm GeV}}
\newcommand{\GeVx}{\rm GeV}
\newcommand{\gevx}{\rm GeV}
\newcommand{\GeVc}{\rm GeV/c}
\newcommand{\gevc}{\rm GeV/c}
\newcommand{\MeVc}{\rm MeV/c}
\newcommand{\mevc}{\rm MeV/c}
\newcommand{\MeV}{\mbox{\rm MeV}}
\newcommand{\mev}{\mbox{\rm MeV}}
\newcommand{\MeVx}{\mbox{\rm MeV}}
\newcommand{\mevx}{\mbox{\rm MeV}}
\newcommand{\GeVsq}{\mbox{${\rm GeV}^2$}}
\newcommand{\gevsq}{\mbox{${\rm GeV}^2$}}
\newcommand{\gevsqc}{\mbox{${\rm GeV^2/c^4}$}}
\newcommand{\gevcsq}{\mbox{${\rm GeV/c^2}$}}
\newcommand{\mevcsq}{\mbox{${\rm MeV/c^2}$}}
\newcommand{\GeVsqm}{\mbox{${\rm GeV}^{-2}$}}
\newcommand{\gevsqm}{\mbox{${\rm GeV}^{-2}$}}
\newcommand{\nb}{\mbox{${\rm nb}$}}
\newcommand{\nbinv}{\mbox{${\rm nb^{-1}}$}}
\newcommand{\pbinv}{\mbox{${\rm pb^{-1}}$}}
\newcommand{\mm}{\mbox{$\cdot 10^{-2}$}}
\newcommand{\mmm}{\mbox{$\cdot 10^{-3}$}}
\newcommand{\mmmm}{\mbox{$\cdot 10^{-4}$}}
\newcommand{\degr}{\mbox{$^{\circ}$}}
%
%
\newcommand{\F}{$ F_{2}(x,Q^2)\,$}  
\newcommand{\Fc}{$ F_{2}\,$}    
\newcommand{\XP}{x_{{I\!\!P}/{p}}}       
\newcommand{\TOSS}{x_{{i}/{\PO}}}        
\newcommand{\un}[1]{\mbox{\rm #1}} 
\newcommand{\LO}{Leading Order}
\newcommand{\NLO}{Next to Leading Order}
\newcommand{\ft}{$ F_{2}\,$}
%
%
\newcommand{\mc}{\multicolumn}
\newcommand{\bce}{\begin{center}}
\newcommand{\ece}{\end{center}}
\newcommand{\beq}{\begin{equation}}
\newcommand{\eeq}{\end{equation}}
\newcommand{\bea}{\begin{eqnarray}}
\newcommand{\eea}{\end{eqnarray}}
%
%
\def\lsim{\mathrel{\rlap{\lower4pt\hbox{\hskip1pt$\sim$}}
    \raise1pt\hbox{$<$}}}         
\def\gsim{\mathrel{\rlap{\lower4pt\hbox{\hskip1pt$\sim$}}
    \raise1pt\hbox{$>$}}}         
%
%
\newcommand{\pom}{{I\!\!P}}
\newcommand{\PO}{I\!\!P}
\newcommand{\slowpi}{\pi_{\mathit{slow}}}
\newcommand{\fiidiii}{F_2^{D(3)}}
\newcommand{\fiidiiiarg}{\fiidiii\,(\beta,\,Q^2,\,x)}
\newcommand{\n}{1.19\pm 0.06 (stat.) \pm0.07 (syst.)}
\newcommand{\nz}{1.30\pm 0.08 (stat.)^{+0.08}_{-0.14} (syst.)}
\newcommand{\fiidiiiful}{F_2^{D(4)}\,(\beta,\,Q^2,\,x,\,t)}
\newcommand{\fiipom}{\tilde F_2^D}
\newcommand{\ALPHA}{1.10\pm0.03 (stat.) \pm0.04 (syst.)}
\newcommand{\ALPHAZ}{1.15\pm0.04 (stat.)^{+0.04}_{-0.07} (syst.)}
\newcommand{\fiipomarg}{\fiipom\,(\beta,\,Q^2)}
\newcommand{\pomflux}{f_{\pom / p}}
\newcommand{\nxpom}{1.19\pm 0.06 (stat.) \pm0.07 (syst.)}
\newcommand {\gapprox}
   {\raisebox{-0.7ex}{$\stackrel {\textstyle>}{\sim}$}}
\newcommand {\lapprox}
   {\raisebox{-0.7ex}{$\stackrel {\textstyle<}{\sim}$}}
\newcommand{\pomfluxarg}{f_{\pom / p}\,(x_\pom)}
\newcommand{\dsf}{\mbox{$F_2^{D(3)}$}}
\newcommand{\dsfva}{\mbox{$F_2^{D(3)}(\beta,Q^2,x_{I\!\!P})$}}
\newcommand{\dsfvb}{\mbox{$F_2^{D(3)}(\beta,Q^2,x)$}}
\newcommand{\dsfpom}{$F_2^{I\!\!P}$}
\newcommand{\gap}{\stackrel{>}{\sim}}
\newcommand{\lap}{\stackrel{<}{\sim}}
\newcommand{\fem}{$F_2^{em}$}
\newcommand{\tsnmp}{$\tilde{\sigma}_{NC}(e^{\mp})$}
\newcommand{\tsnm}{$\tilde{\sigma}_{NC}(e^-)$}
\newcommand{\tsnp}{$\tilde{\sigma}_{NC}(e^+)$}
\newcommand{\st}{$\star$}
\newcommand{\sst}{$\star \star$}
\newcommand{\ssst}{$\star \star \star$}
\newcommand{\sssst}{$\star \star \star \star$}
\newcommand{\tw}{\theta_W}
\newcommand{\sw}{\sin{\theta_W}}
\newcommand{\cw}{\cos{\theta_W}}
\newcommand{\sww}{\sin^2{\theta_W}}
\newcommand{\cww}{\cos^2{\theta_W}}
\newcommand{\trm}{m_{\perp}}
\newcommand{\trp}{p_{\perp}}
\newcommand{\trmm}{m_{\perp}^2}
\newcommand{\trpp}{p_{\perp}^2}
\newcommand{\ev}{\'ev\'enement}
\newcommand{\evs}{\'ev\'enements}
\newcommand{\mdv}{mod\`ele \`a dominance m\'esovectorielle}
\newcommand{\mdmv}{mod\`ele \`a dominance m\'esovectorielle}
\newcommand{\mdm}{mod\`ele \`a dominance m\'esovectorielle}
\newcommand{\idiff}{interaction diffractive}
\newcommand{\idiffs}{interactions diffractives}
\newcommand{\pdmv}{production diffractive de m\'esons vecteurs}
\newcommand{\pdmr}{production diffractive de m\'esons \rh}
\newcommand{\pdmp}{production diffractive de m\'esons \ph}
\newcommand{\pdmo}{production diffractive de m\'esons \om}
\newcommand{\pdm}{production diffractive de m\'esons}
\newcommand{\pdiff}{production diffractive}
\newcommand{\diff}{diffractive}
\newcommand{\produ}{production}
\newcommand{\mv}{m\'eson vecteur}
\newcommand{\mvs}{m\'esons vecteurs}
\newcommand{\me}{m\'eson}
\newcommand{\mr}{m\'eson \rh}
\newcommand{\mph}{m\'eson \ph}
\newcommand{\mo}{m\'eson \om}
\newcommand{\mrs}{m\'esons \rh}
\newcommand{\mps}{m\'esons \ph}
\newcommand{\mos}{m\'esons \om}
\newcommand{\photo}{photoproduction}
\newcommand{\agq}{\`a grand \qsq}
\newcommand{\agqsq}{\`a grand \qsq}
\newcommand{\apq}{\`a petit \qsq}
\newcommand{\apqsq}{\`a petit \qsq}
\newcommand{\de}{d\'etecteur}
%
%
\newcommand{\sqrts}{$\sqrt{s}$}
\newcommand{\Oa}{$O(\alpha_s)$}
\newcommand{\Oaa}{$O(\alpha_s^2)$}
\newcommand{\PT}{p_{\perp}}
\newcommand{\sh}{\hat{s}}
\newcommand{\uh}{\hat{u}}
\newcommand{\ttbs}{\char'134}
\newcommand{\xpomlo}{3\times10^{-4}}
\newcommand{\xpomup}{0.05}
\newcommand{\llq}{$\alpha_s \ln{(\qsq / \Lambda_{QCD}^2)}$}
\newcommand{\llqx}{$\alpha_s \ln{(\qsq / \Lambda_{QCD}^2)} \ln{(1/x)}$}
\newcommand{\llx}{$\alpha_s \ln{(1/x)}$}
%
%
\newcommand{\Brodsky}{Brodsky {\it et al.}}
\newcommand{\FKS}{Frankfurt, Koepf and Strikman}
\newcommand{\Kop}{Kopeliovich {\it et al.}}
\newcommand{\Ginzburg}{Ginzburg {\it et al.}}
\newcommand{\Ryskin}{\mbox{Ryskin}}
\newcommand{\Kaidalov}{Kaidalov {\it et al.}}
%
%
\def\ar#1#2#3   {{\em Ann. Rev. Nucl. Part. Sci.} {\bf#1} (#2) #3}
\def\epj#1#2#3  {{\em Eur. Phys. J.} {\bf#1} (#2) #3}
\def\err#1#2#3  {{\it Erratum} {\bf#1} (#2) #3}
\def\ib#1#2#3   {{\it ibid.} {\bf#1} (#2) #3}
\def\ijmp#1#2#3 {{\em Int. J. Mod. Phys.} {\bf#1} (#2) #3}
\def\jetp#1#2#3 {{\em JETP Lett.} {\bf#1} (#2) #3}
\def\mpl#1#2#3  {{\em Mod. Phys. Lett.} {\bf#1} (#2) #3}
\def\nim#1#2#3  {{\em Nucl. Instr. Meth.} {\bf#1} (#2) #3}
\def\nc#1#2#3   {{\em Nuovo Cim.} {\bf#1} (#2) #3}
\def\np#1#2#3   {{\em Nucl. Phys.} {\bf#1} (#2) #3}
\def\pl#1#2#3   {{\em Phys. Lett.} {\bf#1} (#2) #3}
\def\prep#1#2#3 {{\em Phys. Rep.} {\bf#1} (#2) #3}
\def\prev#1#2#3 {{\em Phys. Rev.} {\bf#1} (#2) #3}
\def\prl#1#2#3  {{\em Phys. Rev. Lett.} {\bf#1} (#2) #3}
\def\ptp#1#2#3  {{\em Prog. Th. Phys.} {\bf#1} (#2) #3}
\def\rmp#1#2#3  {{\em Rev. Mod. Phys.} {\bf#1} (#2) #3}
\def\rpp#1#2#3  {{\em Rep. Prog. Phys.} {\bf#1} (#2) #3}
\def\sjnp#1#2#3 {{\em Sov. J. Nucl. Phys.} {\bf#1} (#2) #3}
\def\spj#1#2#3  {{\em Sov. Phys. JEPT} {\bf#1} (#2) #3}
\def\zp#1#2#3   {{\em Zeit. Phys.} {\bf#1} (#2) #3}
%
%
\newcommand{\clearemptydoublepage}{\newpage{\pagestyle{empty}\cleardoublepage}}
\newcommand{\scaption}[1]{\caption{\protect{\footnotesize  #1}}}
\newcommand{\proc}[2]{\mbox{$ #1 \rightarrow #2 $}}
\newcommand{\average}[1]{\mbox{$ \langle #1 \rangle $}}
\newcommand{\av}[1]{\mbox{$ \langle #1 \rangle $}}

\vspace*{1.0cm}
\begin{center}
\boldmath
{\bf \LARGE Elastic production of Vector Mesons at HERA:} \\ [3mm]
{\bf \LARGE  study of the scale of the interaction and } \\ [2mm]
{\bf \LARGE  measurement of the helicity amplitudes~\footnote{
  Plots presented in the vector meson
  discussion session of the Workshop of Low-$x$ Physics (June 1999) 
  in Tel-Aviv, Israel.} } \\ [3mm]
\unboldmath
\end{center}
\vspace*{0.6cm}
\begin{center}
{Barbara CLERBAUX}\\ 
{Universit\'e Libre de Bruxelles}\\ 
{e-mail: barbara.clerbaux@hep.iihe.ac.be}\\ 
\end{center}
%
\begin{center}
{\bf Abstract}
\begin{quotation}
\noindent A compilation of H1 and ZEUS cross section measurements for 
elastic vector meson production is presented as a function of the scale
$K^2 = (Q^2+M_V^2)/4$, where \qsq\ is the exchanged photon virtuality and
$M_V$ is the mass of the vector meson. 
The ratio of longitudinal to transverse cross sections 
$R$ = $\sigma_L/\sigma_T$ is presented as a function of $Q^2/M^2_V$.
The cross sections are separated in
a transverse and a longitudinal component and are presented as a function of the scale
$K^2$. The intercept $\alpha(0)$ -- 1 governing the energy dependence of the vector meson
cross sections is compared with the $\lambda$ parameter measured in inclusive 
$F_2$ analysis. 
For \rh\ meson production, the helicity amplitude ratios 
$|T_{ij}|/|T_{00}|$, extracted from the H1 and ZEUS measurements of the
spin density matrix elements are presented as a function of \qsq\ and \W\ and are 
compared to recent predictions.
\end{quotation}
\end{center}
%
\section{Introduction}

This note includes the most recent HERA results 
on exclusive vector meson (VM) production: $ e + p \rightarrow e + V\!M + p$.
The H1 and ZEUS experiments studied the production of 
\rh~\cite{h1_rho,zeus_rho_gp,zeus_rho_jpsi_hq,zeus_rho_hq_me}, \om~\cite{zeus_ome_gp,zeus_ome_hq},
\ph~\cite{h1_phi,zeus_phi_gp,zeus_phi_hq}, \jpsi~\cite{h1_jpsi_gp,h1_jpsi_hq,zeus_jpsi_gp,zeus_rho_jpsi_hq} and 
$\Upsilon$~\cite{h1_upsilon,zeus_upsilon} mesons, 
in a \qsq\ domain ranging from photoproduction (\qsq\ $\simeq$ 0) to
\qsq\ = 60 \gevsq\ (where \qsq\ is the virtuality of the photon exchanged
in the interaction).

The note is divided in two parts. The first part presents the
\qsq\ and \w\ dependences of the total, transverse 
and longitudinal cross sections, for the various vector mesons.
The second part is concerned only with \rh\ meson production, 
and presents measurements of the helicity amplitude ratios $|T_{ij}|/|T_{00}|$.

\boldmath
\section{The cross section $\sigma ( \gamma^{(*)} p \rightarrow V\! M \ p)$ }
\unboldmath

\subsection{Scale of the interaction}

Recent measurements of \rh\ meson electroproduction at \qsq\ 
$\gsim$ 10 \gevsq\ ~\cite{h1_rho,zeus_rho_jpsi_hq} and of \jpsi\ meson photo-- and
electroproduction~\cite{h1_jpsi_gp,h1_jpsi_hq,zeus_rho_jpsi_hq}
indicate a strong energy dependence of the $\gamma^{(*)}p \rightarrow V\!M \ p$
cross sections (``hard" behaviour):  $\sigma \propto W^{\delta}$ with $\delta$ $\simeq$ 0.8,
where $W$ is the energy in the photon--proton centre of mass.
This behaviour indicates that the mass of the $c$ quark or a high \qsq\ 
value provides a ``scale" $K^2$ in the
interaction. In this section, results are 
presented for all VM production in a global way
as a function of the scale $K^2 = (Q^2 + M^2_V)/4$, where $M_V$ is the
VM mass. This approach is inspired e.g. by the discussions in reference~\cite{scale1}. 

\begin{figure}[btp]
\setlength{\unitlength}{1.0cm}
\begin{center}
\begin{picture}(16.0,16.0)
\put(0.0,0.){\epsfig{file=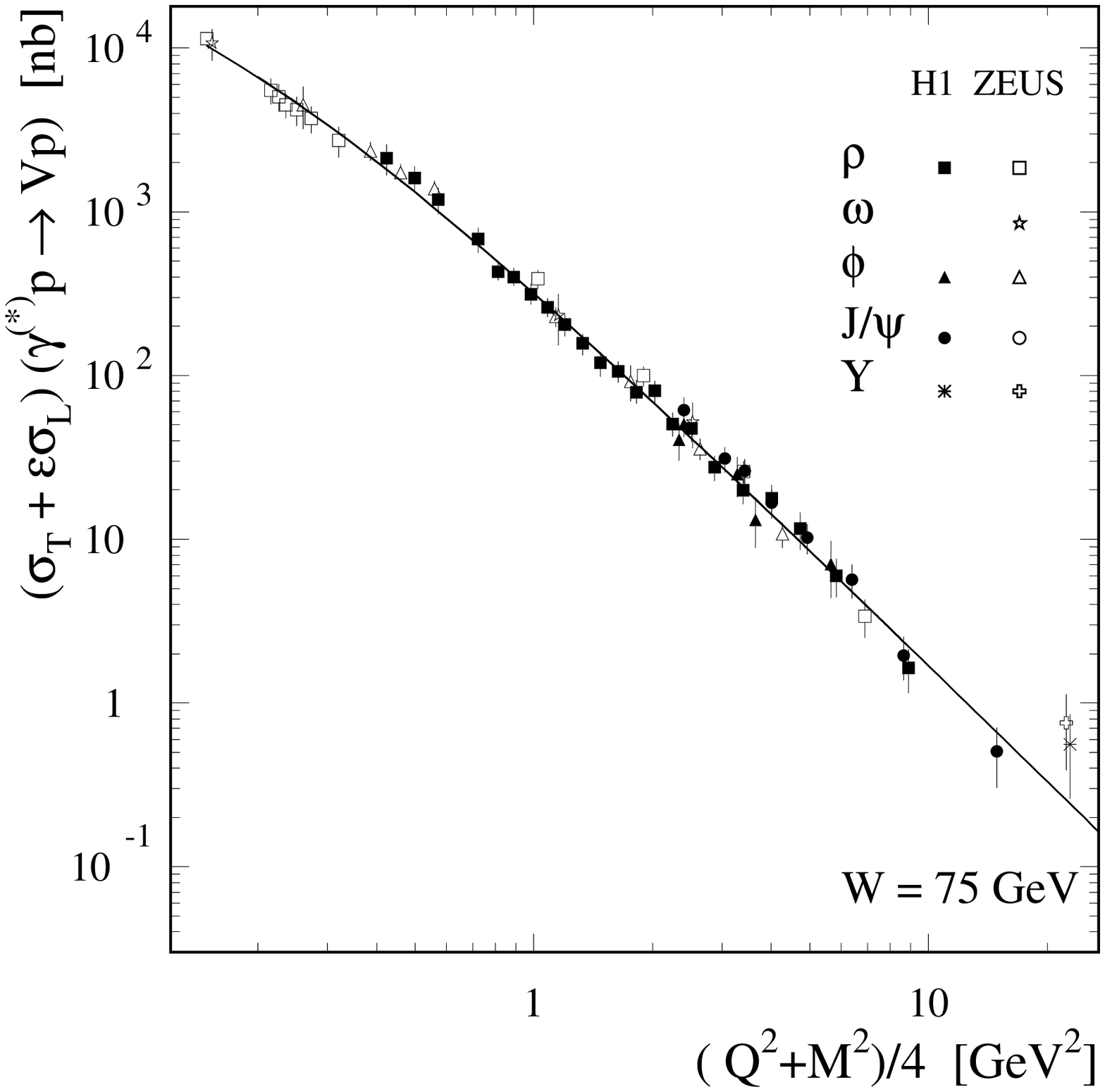,width=16cm,height=16cm}}
\put(2.5,2.4){\epsfig{file=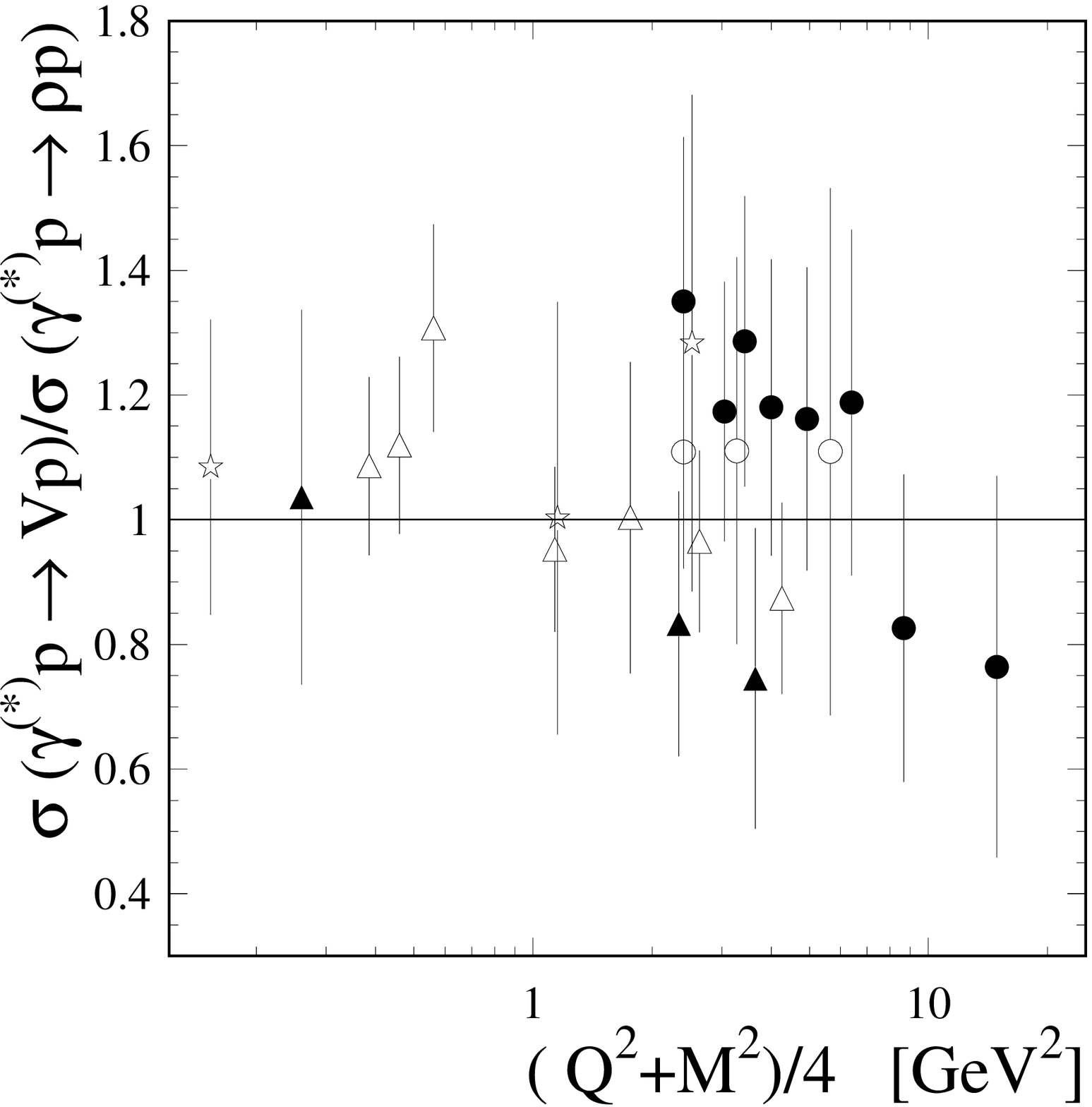,width=7.5cm,height=6cm}}
\end{picture}
    \caption{H1 and ZEUS measurements of the total cross sections
     $\sigma (\gp^{(*)} \rightarrow V\!M \ p)$ as a function of (\qsq\ + $M^2_V$)/4 
     for \rh, \om, \ph, \jpsi\ and $\Upsilon$ vector meson elastic production, at 
     the fixed value of $W$ = 75 GeV~\protect\cite{h1_rho,zeus_rho_gp,zeus_rho_jpsi_hq,zeus_rho_hq_me,zeus_ome_gp,zeus_ome_hq,h1_phi,zeus_phi_gp,zeus_phi_hq,h1_jpsi_gp,h1_jpsi_hq,zeus_jpsi_gp,h1_upsilon,zeus_upsilon}.
     The error bars represent the total errors. The curve corresponds to a fit to 
     the H1 and ZEUS \rh\ data.}
\label{fig:q2_dep}
\end{center}
\end{figure}

A compilation of the HERA measurement of the $\gamma^{(*)}p \rightarrow V\!M \ p$
cross sections is presented in Fig.~\ref{fig:q2_dep} as a function of 
$(Q^2 + M^2_V)/4$. 
The cross sections were scaled by SU(4) factors, according to the 
quark content of the VM : $*$9/1 for the \om, $*$9/2 for the \ph,
$*$9/8 for the \jpsi\ and $*$9/2 for the $\Upsilon$ mesons.
In Fig.~\ref{fig:q2_dep} and in all following plots, the errors on the data points
represent the full errors (including the statistical,
systematic and normalisation errors, added in quadrature).

The cross sections are measured at \W\ = 75 GeV, or are moved to this value
according to the parametrisation $\sigma \propto W^\delta$, using the 
$\delta$ value measured by the relevant experiment.
The ZEUS \rh\ and \ph\ cross sections were corrected for the
$|t|$ cut ($|t|$ $<$ 0.5 or 0.6 \gevsq), following an exponentially 
falling distribution d$\sigma/dt \propto e^{bt}$, with a \qsq\ dependent
$b$ parameter according to measurements (this correction is $\lsim$ 7 \%).

Within experimental errors, the total cross sections for VM production,
including the SU(4) normalisation factors, appear to lay on a
universal curve when plotted as a function of the scale $K^2 = (Q^2 + M^2_V)/4$,
except possibly for the $\Upsilon$ photoproduction~\footnote
   {The cross sections $\sigma (\gamma p \rightarrow \Upsilon({\rm 1S}) p)$ measured by
    H1 and ZEUS at \W\ = 160 and 120 GeV respectively, were moved to the value
    \W\ = 75 GeV using the parametrisation $\sigma \propto W^\delta$, with $\delta$ = 1.7. 
    This high value of the parameter $\delta$ comes from the prediction 
    of~\protect\cite{martin}. Note that if the value $\delta$ = 0.8 is used (a value measured
    in case of \jpsi\ photoproduction), the cross sections are higher 
    by a factor 1.5 for ZEUS and 2.0 for H1.}.

A fit performed on the H1 and ZEUS \rh\ data using the parametrisation 
$\sigma = a (K^2 + b^2)^c$, with $b^2$ = 0.11 $\pm$ 0.03 \gevsq\ and 
$c$ = -- 2.37 $\pm$ 0.10 ($\chi^2/ndf$ = 0.67) is shown 
as the full curve in Fig.~\ref{fig:q2_dep}.
The ratio of the \om, \ph\ and \jpsi\ cross sections to this 
parametrisation is presented in the lower plot of Fig.~\ref{fig:q2_dep}.

It is interesting to recognise that the universal $(Q^2 + M^2_V)/4$ dependence
is for the total cross section measurements.

\subsection{Ratio of the longitudinal to transverse cross sections}

\begin{figure}[bth]
\setlength{\unitlength}{1.0cm}
\begin{center}
\begin{picture}(14.0,12.0)
\put(0.0,0.0){\epsfig{file=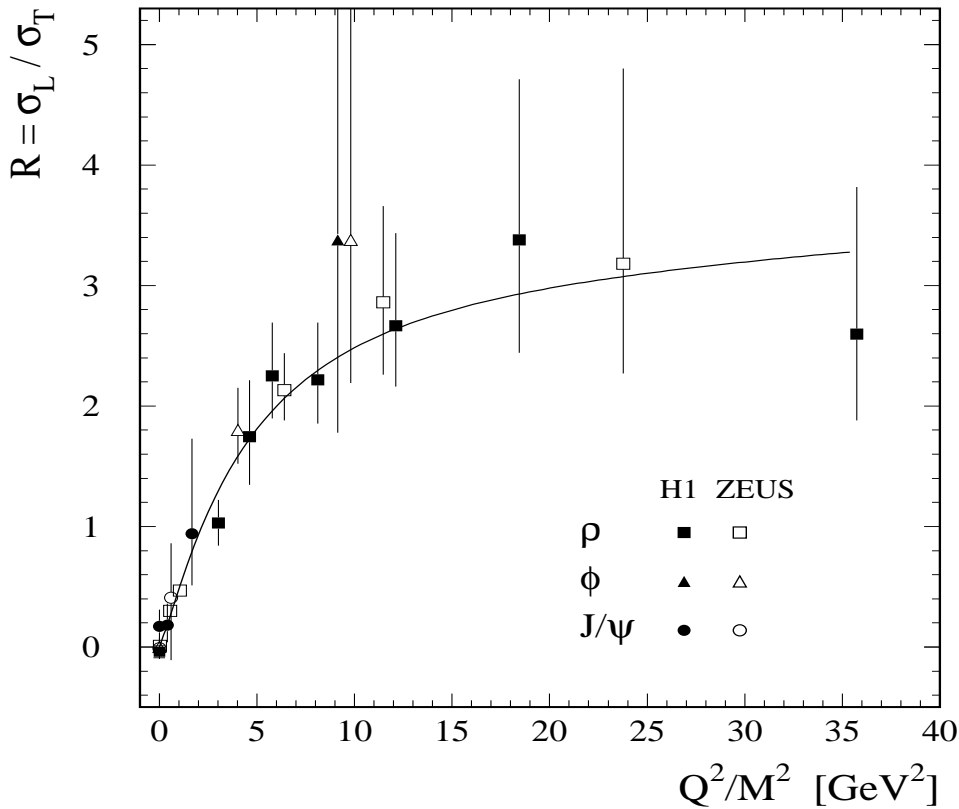,width=14cm,height=12cm}}
\end{picture}
     \caption{H1 and ZEUS measurements of $R$ = $\sigma_L / \sigma_T$
      as a function of $Q^2 / M^2_V$, for \rh, \ph\ and \jpsi\ vector 
      meson elastic production~\protect\cite{h1_rho,zeus_rho_gp,zeus_rho_jpsi_hq,
      zeus_phi_gp,zeus_phi_hq,h1_jpsi_gp,h1_jpsi_hq,zeus_jpsi_gp}. The error
      bars represent the total errors. The curve corresponds to the fit presented in the
      text.}
\label{fig:r_dep}
\end{center}
\end{figure}

The measured ratio of longitudinal to transverse cross sections $R$ = $\sigma_L/\sigma_T$ 
at \qsq\ = 6 \gevsq\ is approximately 2.5 for \rh\ meson production 
but approximately 0.4 for \jpsi\ meson production. 
However, the ratio $R$ presents a similar dependence for the \rh, \ph\ and \jpsi\
meson production when plotted as a function of $Q^2/M^2_V$ (see Fig.~\ref{fig:r_dep}).
All data are well described by a common empirical parametrisation:
$R =a (Q^2/M^2_V)^b (\ln{Q^2/M^2_V+10})^c$ (curve in Fig.~\ref{fig:r_dep}).

It is observed that $R$ rises steeply at small \qsq, with a weaker dependence
at large \qsq\ values. This behaviour is consistent with the fact that the $Q^2/M^2_V$
dependence expected at leading order is modified by higher order corrections. The data in
Fig.~\ref{fig:r_dep} suggest that, within the present statistical precision, this modification
preserves the ratio $Q^2/M^2_V$ as the relevant variable for $\sigma_L/\sigma_T$.

\subsection{The transverse and longitudinal cross sections}

\begin{figure}[btp]
\setlength{\unitlength}{1.0cm}
\begin{center}
\begin{picture}(13.0,21.0)
\put(0.0,11.0){\epsfig{file=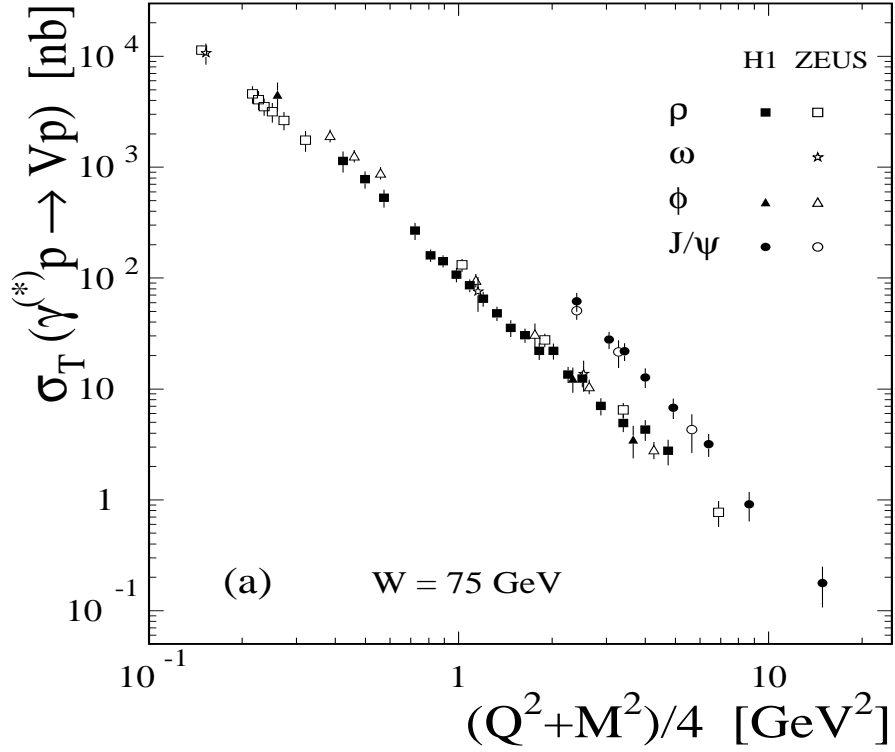,width=13cm,height=11cm}}
\put(0.0,0.0){\epsfig{file=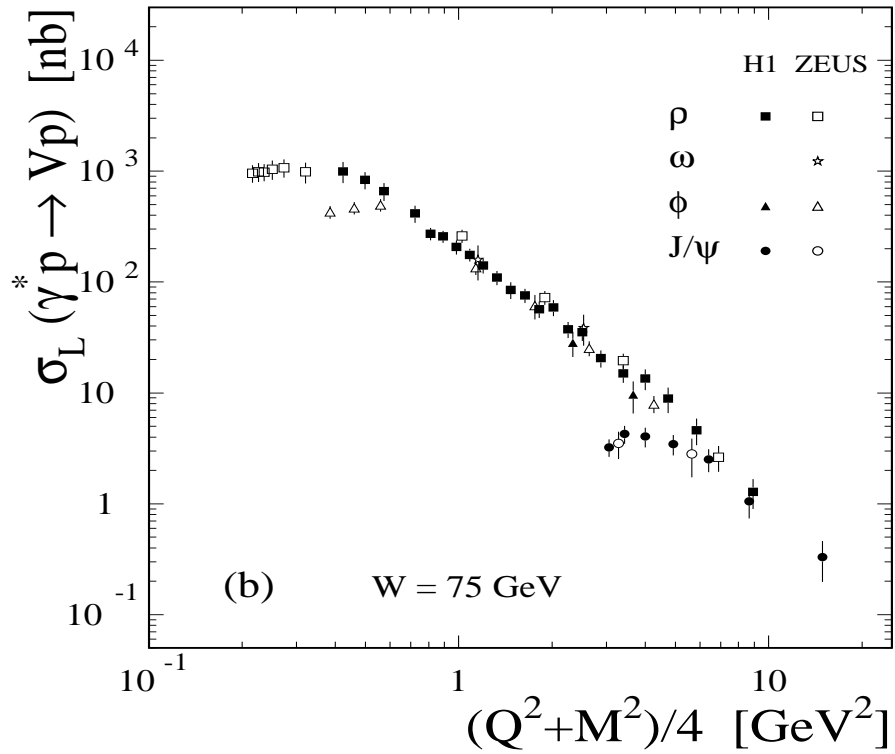,width=13cm,height=11cm}}
\end{picture}
     \caption{H1 and ZEUS measurements of the longitudinal (a) and the
     transverse (b) cross sections as a function of (\qsq\ + $M^2_V$)/4 
     for \rh, \om, \ph\ and \jpsi\ vector meson elastic production, at 
     the fixed value of $W$ = 75 GeV~\protect\cite{h1_rho,zeus_rho_gp,zeus_rho_jpsi_hq,zeus_rho_hq_me,zeus_ome_gp,zeus_ome_hq,h1_phi,zeus_phi_gp,zeus_phi_hq,h1_jpsi_gp,h1_jpsi_hq,zeus_jpsi_gp}.
     The error bars represent the total errors.}
\label{fig:q2_dep_pol}
\end{center}
\end{figure}

Figs.~\ref{fig:q2_dep_pol} (a) and (b) present the transverse 
$\sigma_T$ and the longitudinal $\sigma_L$ cross sections separately,
as a function of $(Q^2 + M^2_V)/4$. 
The parametrisation for $R = \sigma_L / \sigma_T$, described above, was used 
to separate the longitudinal and the transverse parts of the cross sections
for the different vector mesons:
$$\sigma = \sigma_T + \varepsilon \sigma_L = \sigma_T (1 + \varepsilon R), $$
where the polarisation parameter $\av {\varepsilon}$ = 0.996 at HERA.

The transverse cross sections are well described by a simple power law dependence:
$\sigma_T \propto (Q^2+ M^2_V) ^n$, with $n$ = -- 2.47 $\pm$ 0.03 for \rh, 
$n$~=~--~2.4~$\pm$~0.1 for \om, 
$n$~=~--~2.8~$\pm$~0.1 for \ph, and
$n$~=~--~3.1~$\pm$~0.2 for \jpsi\ meson production.

In view of these values of the parameter $n$ ($n \ne 2$), the parametrisation proposed
in the GVDM approach of Schildknecht, Schuler and Surrow~\cite{sss} 
does not provide good fits~\footnote
  {If the transverse and the longitudinal 
   cross sections are fitted using the Schildknecht, Schuler and Surrow parametrisation,
   one obtains for the square of the transverse masses 0.29 $\pm$ 0.01 
   \gevsq\ ($\chi^2/ndf$ = 3.6), 0.39 $\pm$ 0.03 \gevsq\ ($\chi^2/ndf$ = 5.7) 
   and 4.1 $\pm$ 0.4 \gevsq\ ($\chi^2/ndf$ = 1.9) for the \rh, \ph\ and \jpsi\ 
   mesons respectively. The results for the square of the longitudinal masses are 
   0.38 $\pm$ 0.02 \gevsq\ ($\chi^2/ndf$ = 3.1), 0.53 $\pm$ 0.05 \gevsq\ 
   ($\chi^2/ndf$ = 2.2) and 5.4 $\pm$ 0.7 \gevsq\ ($\chi^2/ndf$ = 0.8) 
   for the \rh, \ph\ and \jpsi\ mesons respectively.}.

\subsection{Energy dependence}

The energy dependence of the cross section for VM production at HERA
can be parametrised as $\sigma \propto W^{\delta}$.
In a Regge context, the parameter $\delta$ can be related
to the exchanged trajectory $\alpha (t)$, which is assumed to take a linear form
$\alpha(t) = \alpha(0) + \alpha^\prime \ t \ $, where $t$ is the square of the
four--momentum exchanged at the proton vertex.
The value $\alpha^\prime = 0.25~\gevsqm$ is assumed for the \rh\ and \ph\ mesons, 
as measured in hadron$-$hadron interactions~\cite{dola}. For the \jpsi\ 
meson, the value $\alpha^\prime = 0.05$ is used~\cite{h1_jpsi_gp}.

\begin{figure}[bth]
\setlength{\unitlength}{1.0cm}
\begin{center}
\begin{picture}(14.0,10.0)
\put(0.0,0.0){\epsfig{file=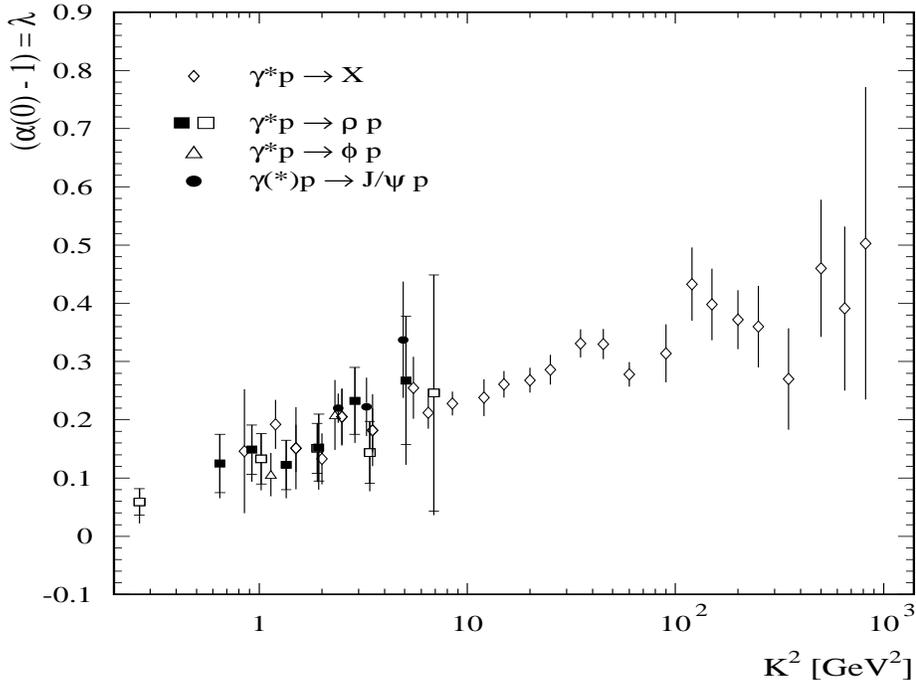,width=14.cm,height=10.0cm}}
\end{picture}
     \caption{The parameter $(\alpha(0) -1)$ and $\lambda$ 
      as a function of the scale $K^2$.
      The H1~\protect\cite{h1_rho,h1_jpsi_gp,h1_jpsi_hq} and 
      ZEUS~\protect\cite{zeus_rho_jpsi_hq,zeus_phi_hq}
      measurements for elastic vector meson production are labelled
      as in Fig.~\protect\ref{fig:q2_dep}. The inclusive $F_2$ measurements from 
      H1~\protect\cite{h1_f2} are labelled using open diamonds.
      The error bars on the data represent the full errors.
      For \rh\ meson production, the outer bars correspond to the case 
      $\alpha^\prime = 0$. The scale is $K^2 = Q^2$ for the $F_2$ inclusive measurement
      and $K^2 = (Q^2+M^2_V)/4$ for elastic vector meson production.}
\label{fig:lambda}
\end{center}
\end{figure}

The values obtained for the parameter ($\alpha(0) -1$)
for the \rh, \ph\ and \jpsi\ meson production are presented in
Fig.~\ref{fig:lambda} as a function of the scale $K^2$ = (\qsq\ + $M^2_V$)/4. 
The error bars on the data represent the full errors.
For the \rh\ meson production, the sensitivity to the choice of $\alpha^\prime$ 
is shown by the outer bars, which contain the variation due to the assumption 
$\alpha^\prime = 0$ (i.e. no shrinkage) added in quadrature.
The points are compared to the values of the parameter $\lambda$ obtained from fits to the
\W\ dependence of inclusive $F_2$ 
measurements~\cite{h1_f2} ($\sigma_{tot} \propto F_2(x,Q^2) \propto
(1/x)^\lambda \propto W^{2\lambda}$), plotted as a function of the scale
$K^2 = Q^2$ (see also~\cite{marco}).

Within experimental errors, a common rise is observed for the
($\alpha(0)-1$) and the $\lambda$ parameters when plotted as a function of the 
relevant scale $K^2$. This is at variance with the values
$1.08 - 1.10$ obtained from fits to
the total and elastic hadron--hadron cross sections~\cite{dola,cudellfit}.

\section{Helicity amplitudes}

This section presents the H1 and ZEUS elastic \rh\ meson production results
on the study of angular distributions of the \rh\ meson production and 
decay~\cite{h1_rho,zeus_rho_hq_me}, which provides information 
on the photon and \rh\ polarisation.

Following the formalism of~\cite{shilling-wolff}, the normalised angular decay 
distribution is a function of 15 combinations of spin density matrix elements,
$r^{\alpha\beta}_{ij}$.
Each of these $r^{\alpha\beta}_{ij}$ is a sum of bilinear combinations 
of the helicity amplitudes $T_{\lambda_{\rho}\lambda_{\gamma}}$.
One can thus invert the system and extract the helicity amplitudes 
from the measurement of the 15 elements.
The motivations for extracting the helicity amplitudes from the 15 matrix elements 
are the following: \\
1. fundamental quantities are computed, on which $s$-channel helicity 
   conservation~\footnote{
      In case of $s$-channel helicity conservation (SCHC), the helicity
      of the vector meson is equal to that of the photon when the spin quantisation
      axis is taken along the direction of the meson momentum is the $\gamma^*p$
      centre of mass system. In that case, the non-flip helicity amplitudes
      $T_{\lambda_{\rho}\lambda_{\gamma}} = T_{00}$ and $T_{11}$ have non-zero
      values, the single flip ($T_{10}$, $T_{01}$, $T_{0-1}$, $T_{-10}$) and 
      double flip amplitudes ($T_{1-1}$, $T_{-11}$) being zero.} (SCHC)
    hypothesis can be checked directly,\\
2. measurements of the real and imaginary parts of the helicity amplitudes
   can be obtained.\\ 

Only 15 equations are available for 18 unknowns (9 complex helicity amplitudes 
$T_{\lambda_{\rho} \lambda_{\gamma}} = |T_{\lambda_{\rho}
 \lambda_{\gamma}}| e ^{i \varphi_{\lambda_{\rho} \lambda_{\gamma}}}$).
However, both H1~\cite{h1_rho} and ZEUS~\cite{zeus_rho_hq_me} results are compatible with 
natural parity exchange~\footnote{
    Natural parity exchange (NPE) is defined by the following relations between the
    helicity amplitudes: 
    $T_{-\lambda_{\rho} -\lambda_{\gamma}}  = (-1)^{\lambda_{\rho}-\lambda_{\gamma}} \
    T_{\lambda_{\rho} \lambda_{\gamma}}$.} (NPE).
Under this hypothesis, 5 helicity amplitudes remain independent (10 unknown): 
$T_{00}, T_{11}, T_{01}$, $T_{10}$ and $T_{1-1}$.

\boldmath
\subsection{The ratios $|T_{ij}| / |T_{00}|$ }
\unboldmath

Minimum $\chi^2$ fits were performed to measurements of the combinations of the matrix 
element to extract the 5 independent complex helicity amplitudes. 
The normalisation $|T_{00}|$ = 1 and $\varphi_{00}$ = 0 is used. 
In the full fit, the parameters $|T_{1-1}|$, $\varphi_{10}$, $\varphi_{01}$ and $\varphi_{1-1}$ 
were found compatible with zero within large errors.
These parameters are then put to zero, and the remaining free parameters of the fit are 
$|T_{11}|$, $|T_{01}|$, $|T_{10}|$ and $\varphi_{11}$.

\begin{figure}[p]
\setlength{\unitlength}{1.0cm}
\begin{center}
\begin{picture}(16.0,16.0)
\put(0.0,8.0){\epsfig{file=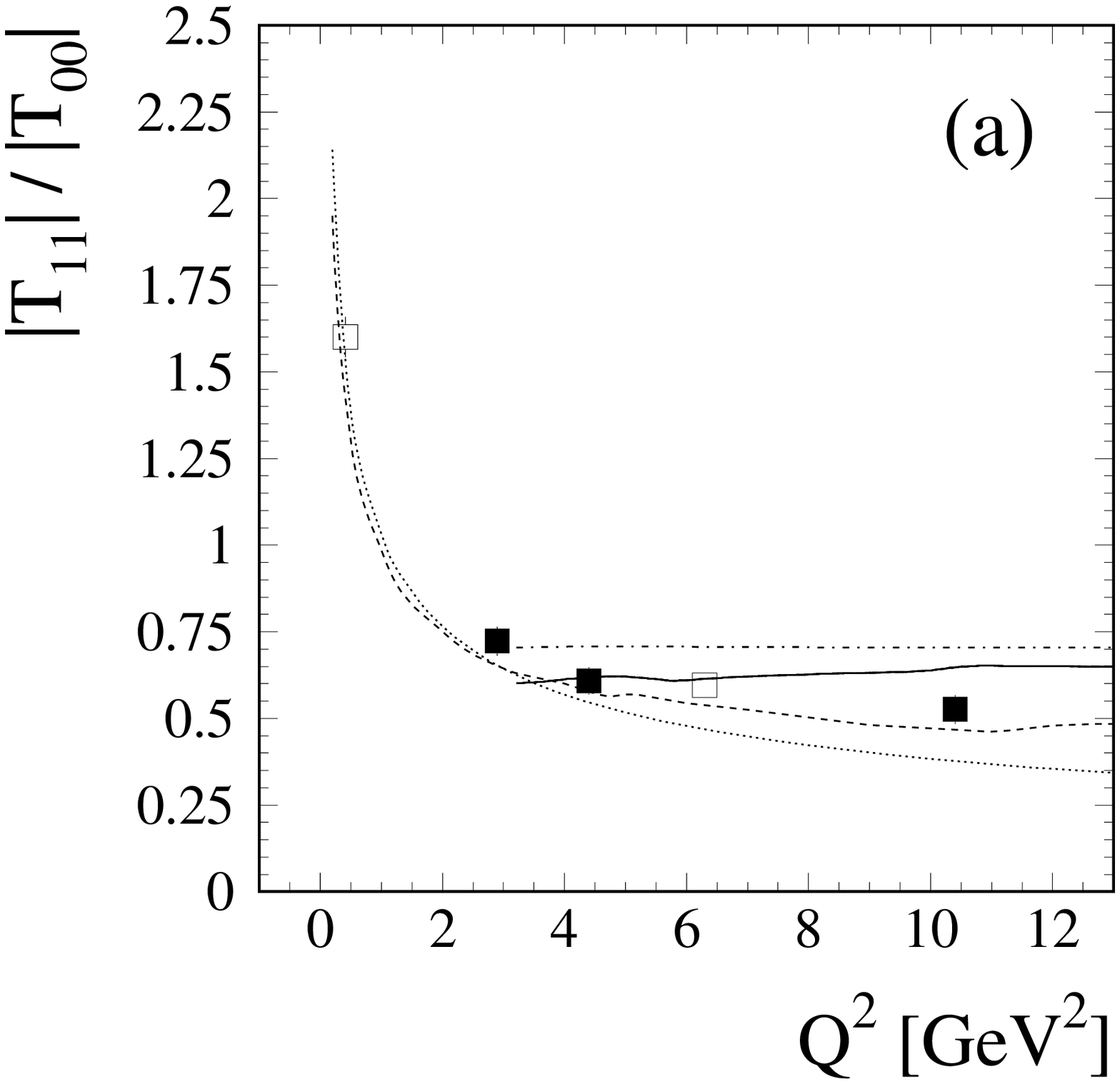,width=8cm,height=8cm}}
\put(8.0,8.0){\epsfig{file=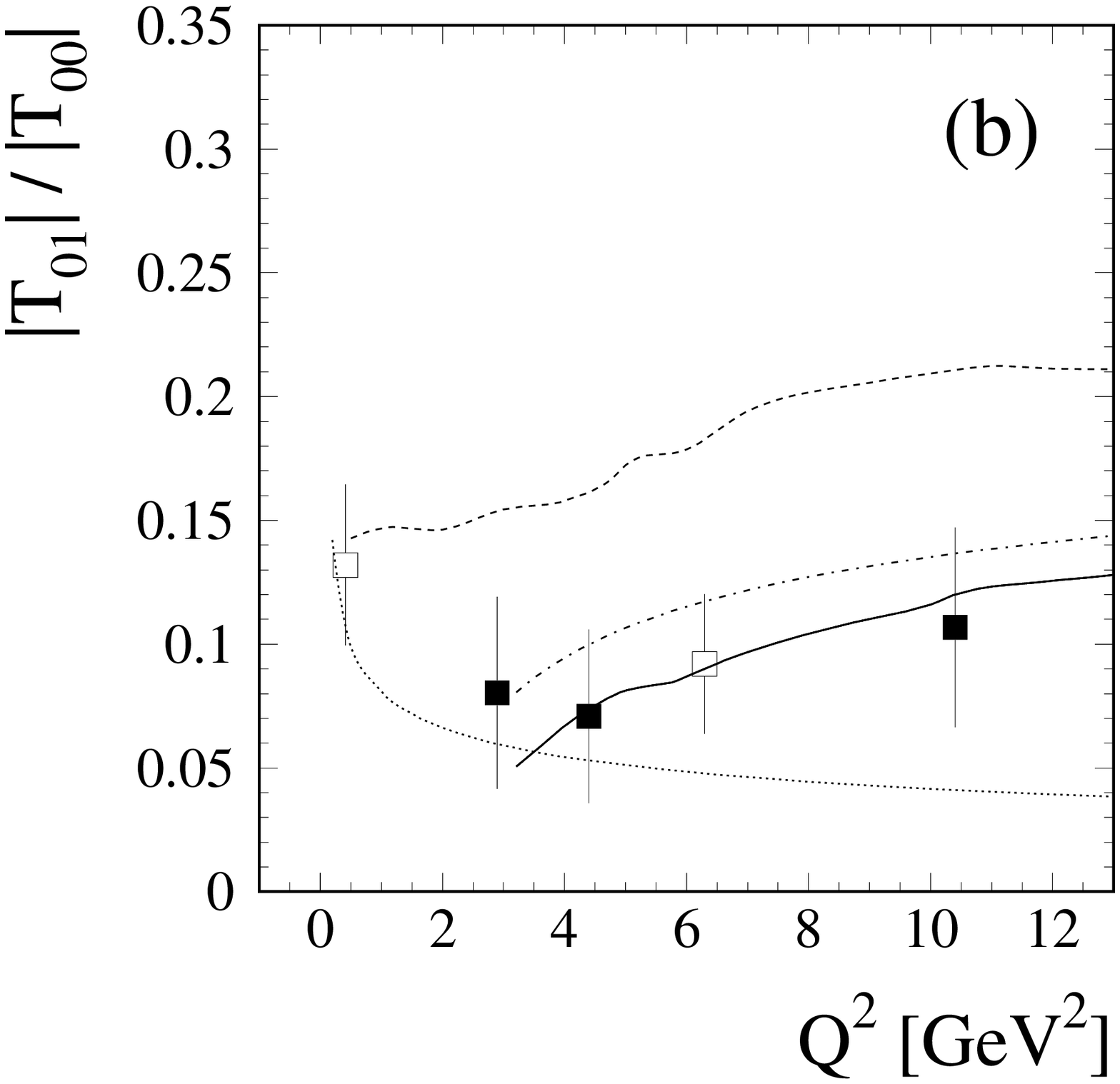,width=8cm,height=8cm}}
\put(0.0,0.0){\epsfig{file=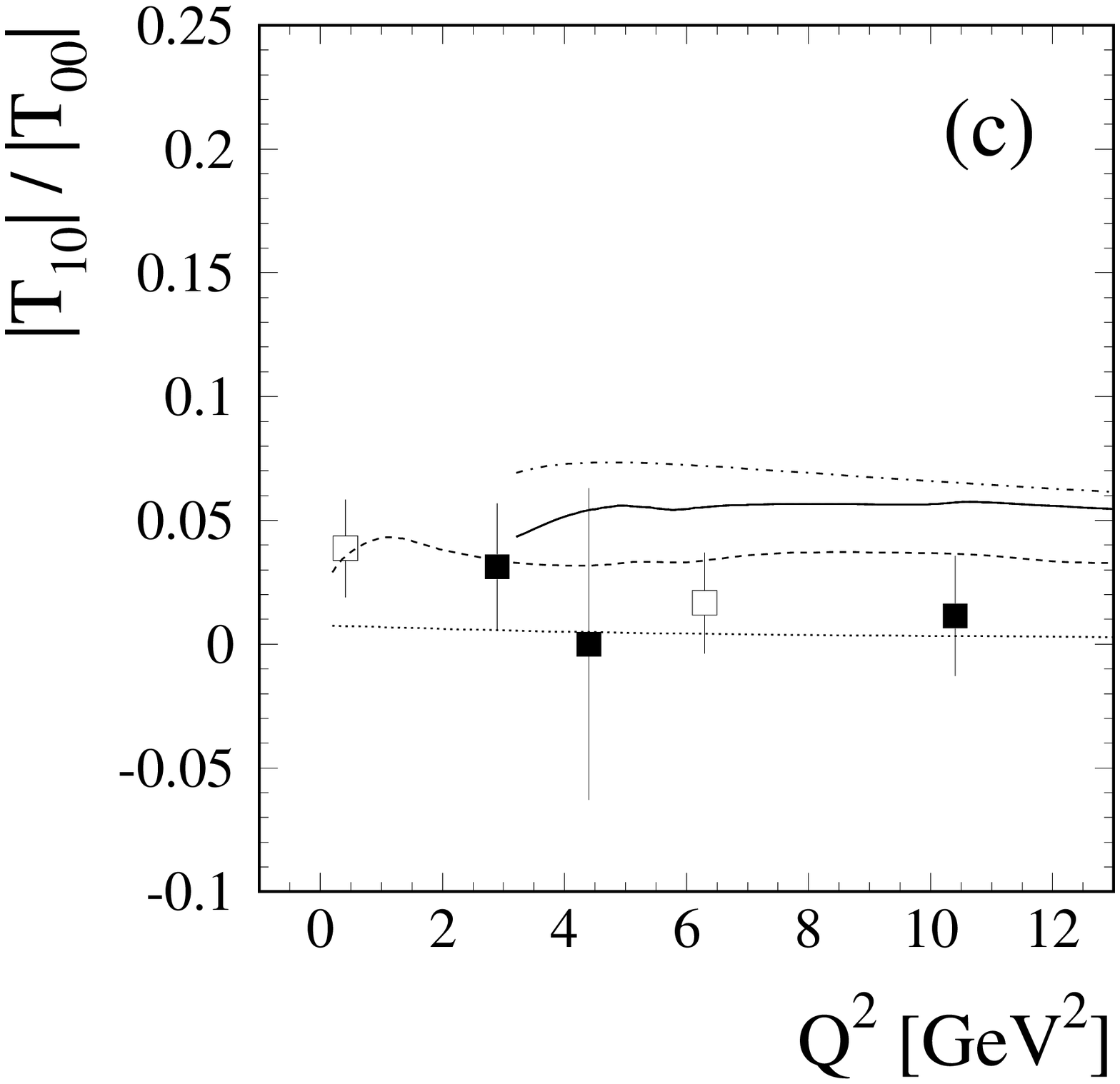,width=8cm,height=8cm}}
\end{picture}
     \caption{\qsq\ dependence of the ratio of the helicity amplitudes: 
      $|T_{11}|/|T_{00}|$ (a), $|T_{01}|/|T_{00}|$ (b) and $|T_{10}|/|T_{00}|$ (c),
      extracted from the H1 and ZEUS measurements of the 15 combinations of spin density
      matrix elements~\protect\cite{h1_rho,zeus_rho_hq_me}.
      The dashed and the dotted lines correspond to predictions of the Royen and Cudell 
      model~\protect\cite{isa} and of the Nikolaev and Akushevich model~\cite{kolya} 
      respectively. The dash-dotted and full lines represent the 
      predictions of the Ivanov and Kirschner model~\protect\cite{ivanov} 
      using respectively the GRV94HO and the CTEQ4lQ parametrisations for the 
      gluon density in the proton.}
\label{fig:ampl}
\end{center}
\end{figure}

The measurement of the ratios $|T_{11}|$ / $|T_{00}|$, $|T_{01}|$ / $|T_{00}|$
and $|T_{10}|$ / $|T_{00}|$ as a function of \qsq\ is presented 
in Figs.~\ref{fig:ampl} (a), (b) and (c) respectively (see also table~\ref{table:val}). 
The H1 data cover the kinematic range 2.5 $<$ \qsq\ $<$ 60 \gevsq, 
30 $<$ W $<$ 140 GeV  ($\av{W}$ = 75 GeV) and $\av {|t|}$ = 0.138 \gevsq. 
For the ZEUS data, the kinematic domain is 0.25 $<$ \qsq\ $<$ 0.85 \gevsq, 
20 $<$ W $<$ 90 GeV  ($\av {W}$ = 45 GeV) and $\av{|t|}$ = 0.14 \gevsq\ for the low
\qsq\ data and 3 $<$ \qsq\ $<$ 30 \gevsq, 40 $<$ W $<$ 120 GeV  ($\av{W}$ = 73 GeV) 
and $\av{|t|}$ = 0.17 \gevsq\ for the high \qsq\ data.

The ratio $|T_{11}|$ / $|T_{00}|$ decreases when \qsq\
increases, indicating that at high \qsq\ longitudinal polarisation dominates.
The ratio $|T_{01}|$ / $|T_{00}|$ is observed to be 
around 8 \% (14.0 $\sigma$ from zero, 5 data points),
and the ratio $|T_{10}|$ / $|T_{00}|$ around 3 -- 4 \% (4.5 $\sigma$ from zero),
which indicate a small but significant violation of SCHC
for \rh\ meson production at HERA.

The predictions of three models based on perturbative QCD are compared 
to the HERA measurements in Fig.~\ref{fig:ampl}.
The dashed lines represent the predictions of the Royen and Cudell model~\cite{isa}, 
computed for $\av{|t|} = 0.135$ \gevsq. The dotted lines are the predictions
of the Nikolaev and Akushevich model~\cite{kolya}, for $\av{|t|} = 0.13$ \gevsq\ and  
$\av{W}$ = 75 GeV.
The dash-dotted and full lines present the predictions of the
Ivanov and Kirschner model~\protect\cite{ivanov}, computed for $\av{|t|} = 0.13$ 
\gevsq\ and $\av{W}$ = 75 GeV, using respectively the GRV94HO and the
CTEQ4lQ parametrisations for the gluon density in the proton. 
These three models predict a violation of SCHC, the amplitudes $T_{01}$ and $T_{10}$
being different from zero, in agreement with the data.

\begin{figure}[p]
\setlength{\unitlength}{1.0cm}
\begin{center}
\begin{picture}(16.0,16.0)
\put(0.0,8.0){\epsfig{file=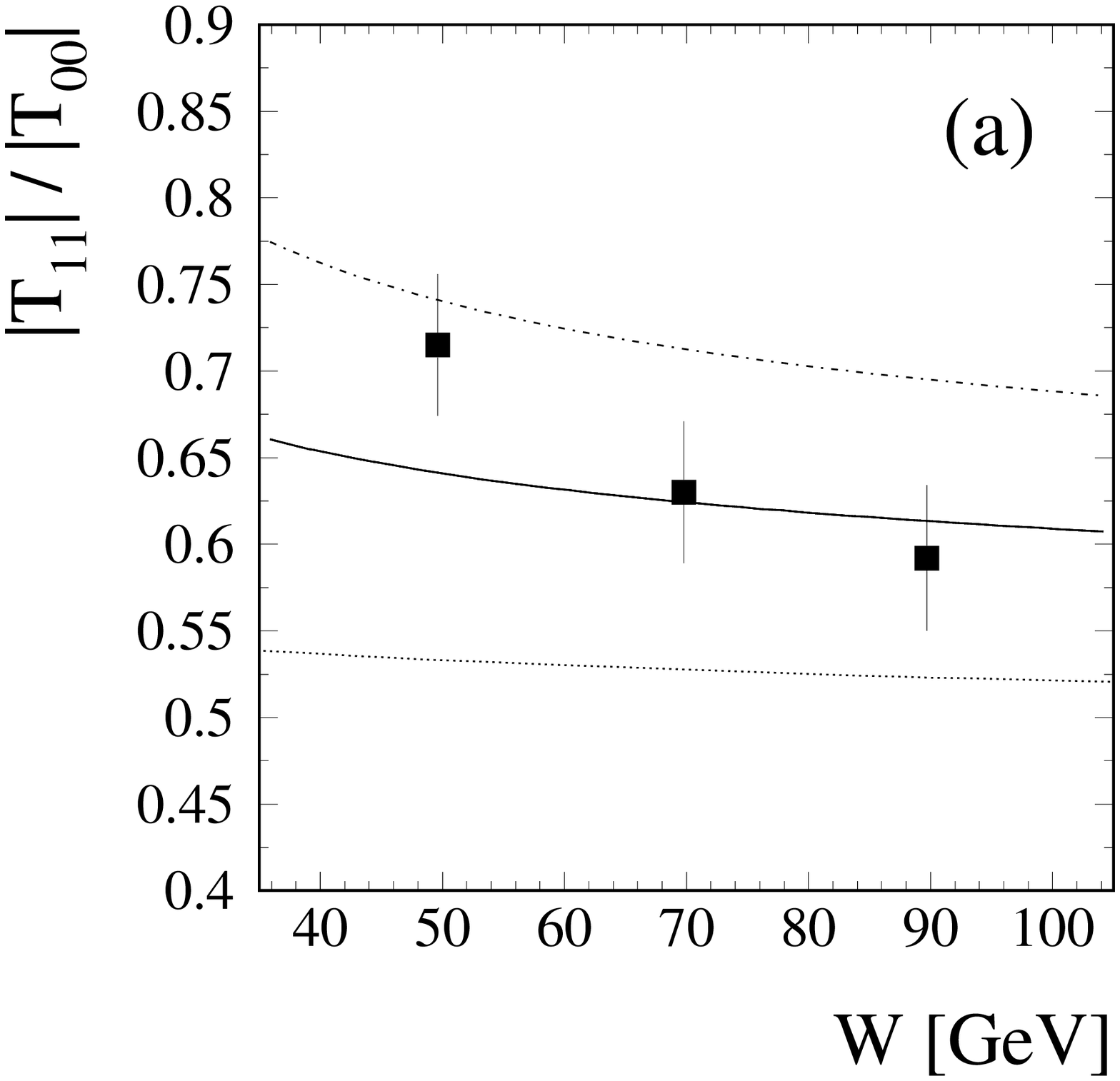,width=8cm,height=8cm}}
\put(8.0,8.0){\epsfig{file=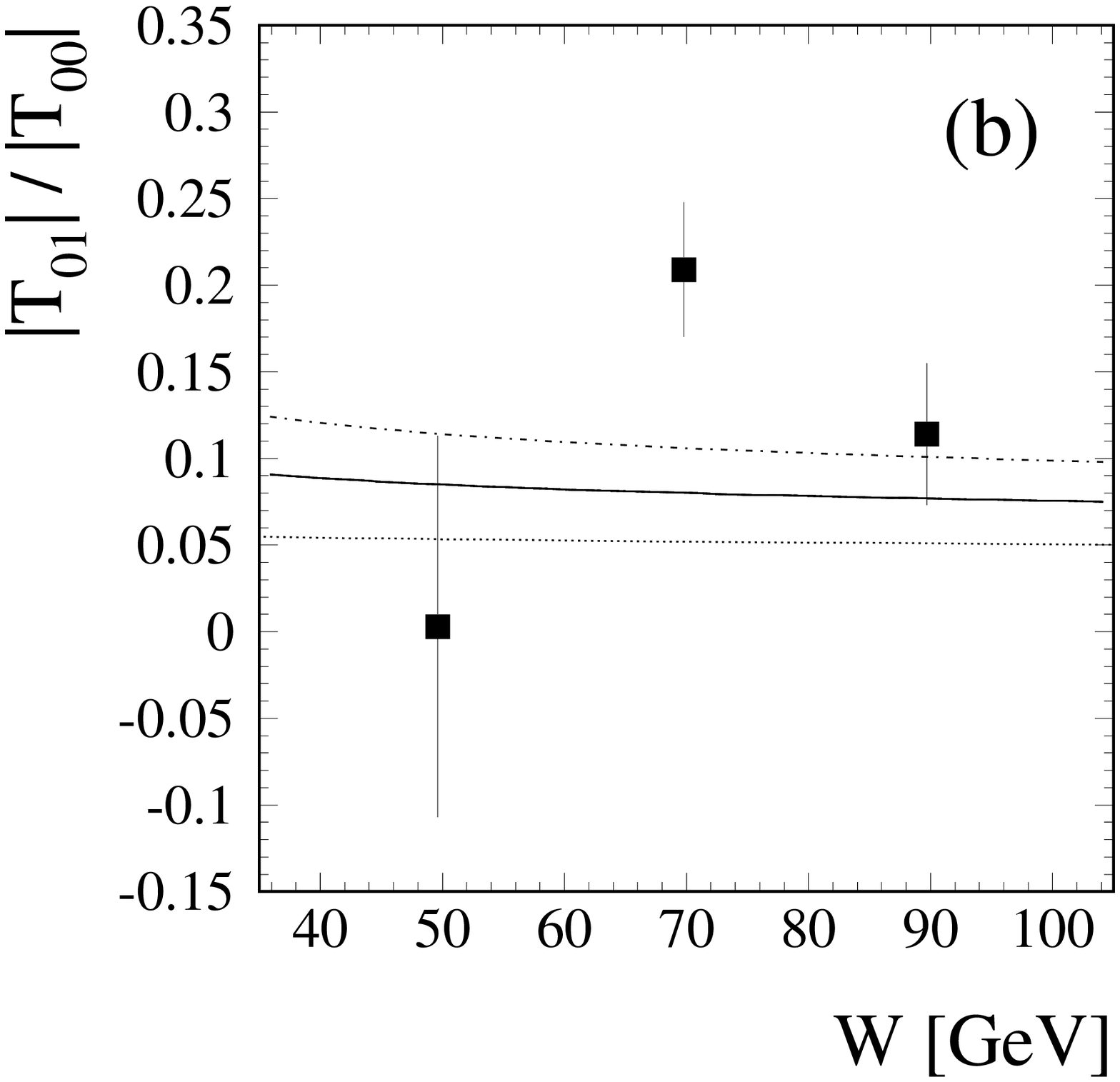,width=8cm,height=8cm}}
\put(0.0,0.0){\epsfig{file=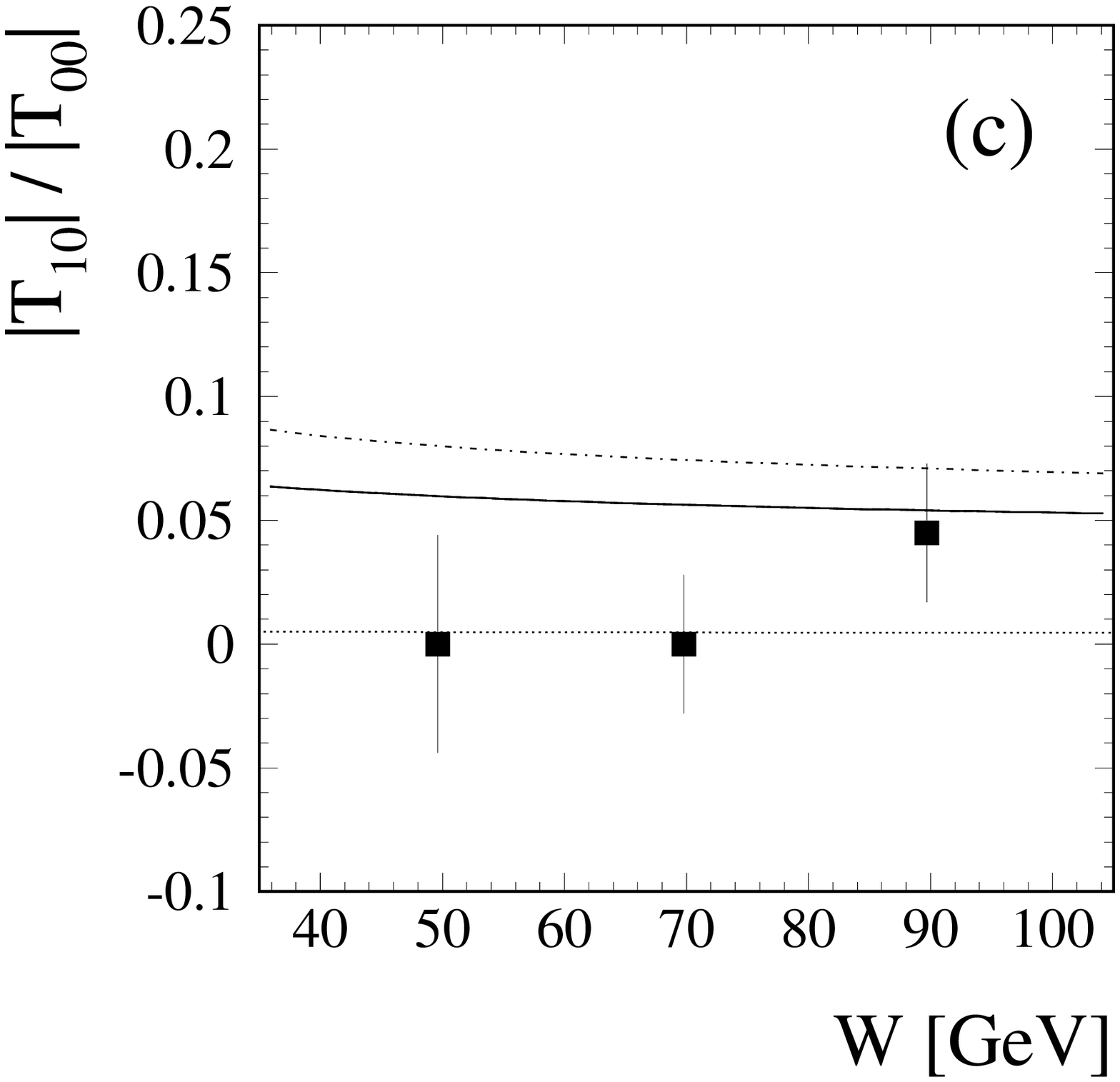,width=8cm,height=8cm}}
\end{picture}
     \caption{\w\ dependence of the ratio of the helicity amplitudes: 
      $|T_{11}|/|T_{00}|$ (a), $|T_{01}|/|T_{00}|$ (b) and $|T_{10}|/|T_{00}|$ (c),
      extracted from the H1 measurements of the 15 combinations of spin density
      matrix elements~\protect\cite{h1_rho}.
      The dotted lines correspond to predictions of the Nikolaev and 
      Akushevich model~\cite{kolya}. The dash-dotted and full lines represent the 
      predictions of the Ivanov and Kirschner model~\protect\cite{ivanov} 
      using respectively the GRV94HO and the CTEQ4lQ parametrisations for the 
      gluon density in the proton.}
\label{fig:ampl_w}
\end{center}
\end{figure}

The \W\ dependence of the ratios $|T_{ij}|$ / $|T_{00}|$ is presented in 
Figs.~\ref{fig:ampl_w} (a), (b) and (c) for $\av{Q^2}$ = 4.8 \gevsq, 
together with predictions of the 
Nikolaev and Akushevich model~\cite{kolya} (dotted lines) and the
Ivanov and Kirschner model~\protect\cite{ivanov} (dash-dotted and full lines) 
(the Royen and Cudell model gives no prediction for the \w\ dependence).


\boldmath
\subsection{The phase $\varphi_{11} - \varphi_{00}$ }
\unboldmath

\begin{figure}[bt]
\setlength{\unitlength}{1.0cm}
\begin{center}
\begin{picture}(14.0,7.0)
\put(0.0,0.0){\epsfig{file=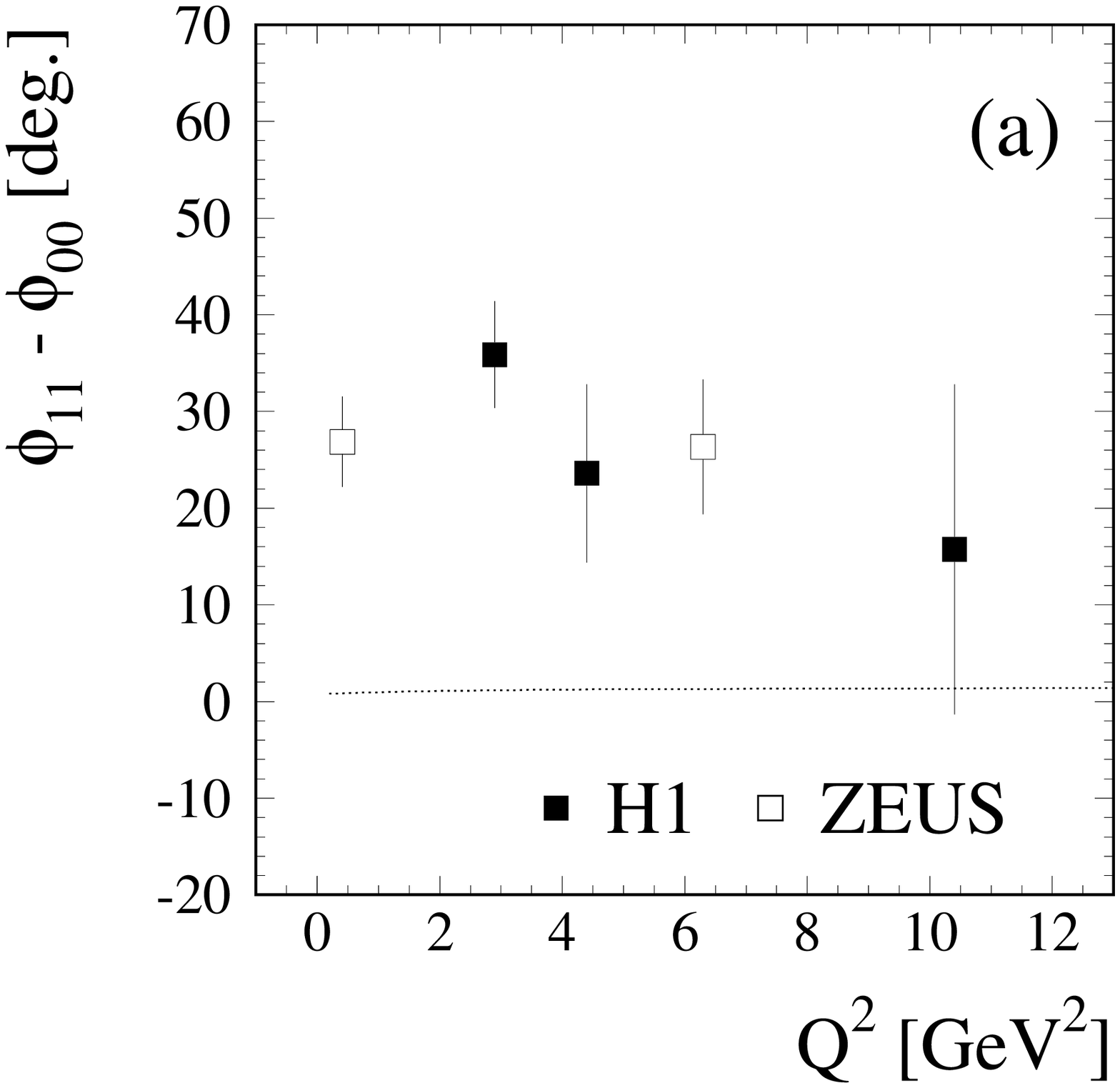,width=7cm,height=7cm}}
\put(7.0,0.0){\epsfig{file=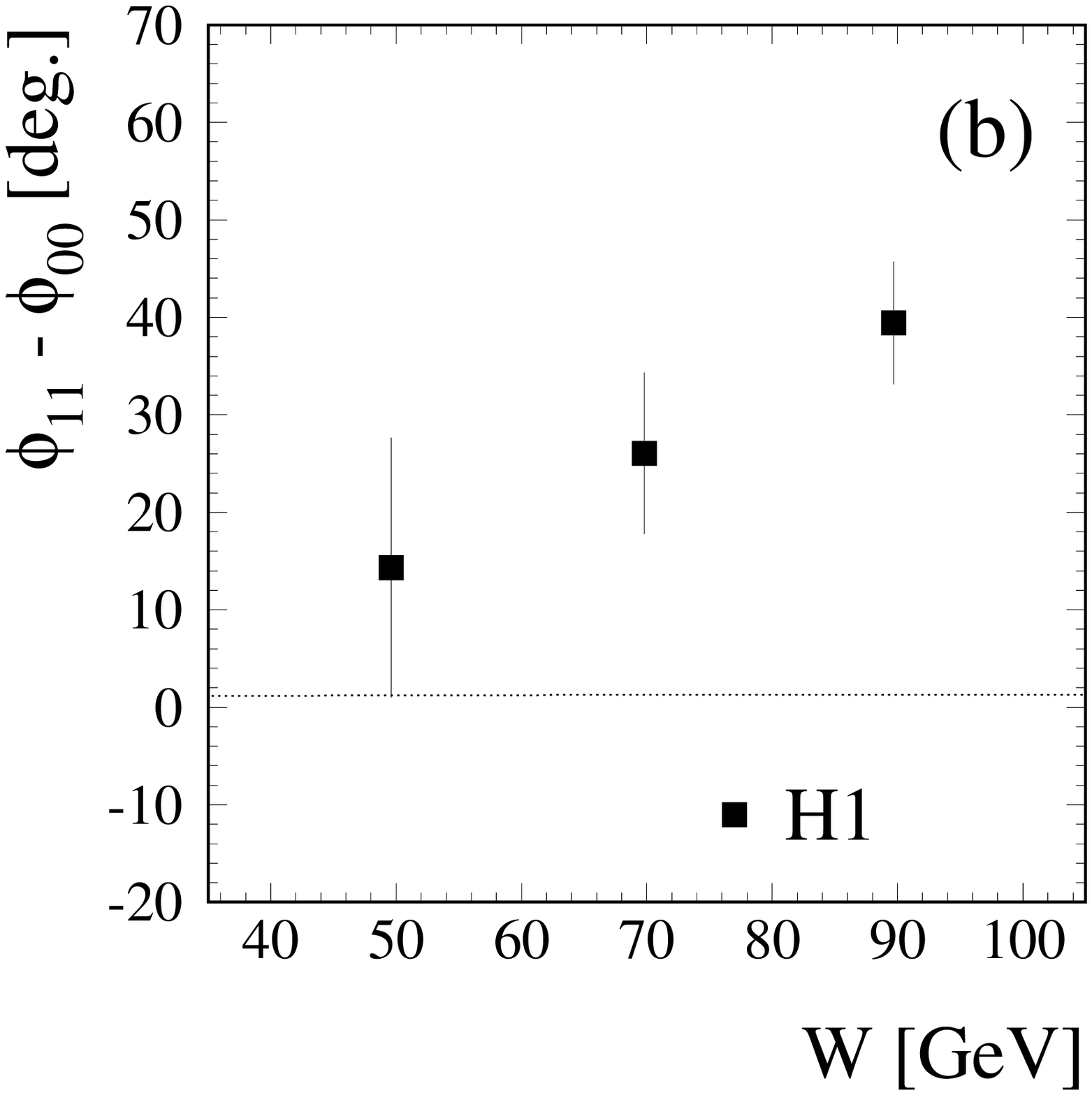,width=7cm,height=7cm}}
\end{picture}
     \caption{ (a) \qsq\  and (b) \w\ dependences of the phase 
      difference $\varphi_{11} - \varphi_{00}$
      extracted from the H1 and ZEUS measurements of the 15 combinations of spin density
      matrix elements~\protect\cite{h1_rho,zeus_rho_hq_me}.
      The dotted lines correspond to predictions of the Nikolaev and 
      Akushevich model~\cite{kolya}.} 
\label{fig:phase}
\end{center}
\end{figure}

Figs.~\ref{fig:phase} (a) and (b) present results for the phase difference 
$\varphi_{11} - \varphi_{00}$, as a function of \qsq\ and \W\ respectively
(see also table~\ref{table:val}).
The phase difference is clearly different from zero, of the order
of $\phi$ = 25 degrees. This value is in agreement with the H1 
measurement~\cite{h1_rho} of 
$\cos{\phi} = 0.925 \pm 0.022$ \mbig{$^{+0.011}_{-0.022}$}
obtained from the angular distribution when supposing SCHC and NPE.
The dotted lines on Figs.~\ref{fig:phase} (a) and (b) represent the
predictions of the Nikolaev and Akushevich model~\cite{kolya}~\footnote{
The models of Royen and Cudell and of Ivanov and Kirschner suppose that
the amplitudes are purely imaginary.}.

\boldmath
\subsection{Fixed target experiments}
\unboldmath

The same procedure was used to extract the helicity amplitudes for the fixed 
target experiments for which the full set of matrix elements have been measured. 
The data in references~\cite{eckardt},~\cite{joos} and~\cite{delpapa} correspond 
to \qsq\ values between 0.4 and 1.1 \gevsq\ and \w\ values between
2.0 and 3.1 \gev, the CHIO experiment~\cite{chio} is at higher energy 
($0 < Q^2 < 3$ \gevsq\ and $12.5 < W < 16$ \gev).
The fitted helicity amplitudes for~\cite{eckardt} are affected by large errors.
Fits to the data in reference~\cite{joos} have a very bad $\chi^2$.
Fits to the data in reference~\cite{delpapa} and to the CHIO data~\cite{chio} 
give helicity amplitude
ratios in agreement, within the experimental errors, with the Royen and Cudell predictions.
The experimental errors on the helicity ratios
for these two fixed target experiments are typically a factor five greater than
the errors on the HERA measurements.

\section*{Acknowledgements}

I am grateful to I. Akushevich, D. Ivanov, P. Marage, N. Nikolaev, I. Royen for 
numerous interesting discussions.

\vspace*{1.5cm}
\begin{table}[h]
    \centering
    \begin{tabular}{|l|c|c|c|c|}
    \hline
    \hline
\qsq\ (\gevsq) & $|T_{11}|/|T_{00}|$ & $|T_{01}|/|T_{00}|$ & $|T_{10}|/|T_{00}|$ 
& $\varphi_{11}-\varphi_{00}$ (deg.) \\
    \hline
0.41& 1.601 $\pm$ 0.058 & 0.132 $\pm$ 0.033 & 0.039 $\pm$ 0.020 & 26.9 $\pm$ 4.7 \\
2.9 & 0.723 $\pm$ 0.041 & 0.080 $\pm$ 0.039 & 0.031 $\pm$ 0.025 & 35.9 $\pm$ 5.5 \\
4.4 & 0.608 $\pm$ 0.039 & 0.071 $\pm$ 0.035 & 0.000 $\pm$ 0.063 & 23.6 $\pm$ 9.2 \\
6.3 & 0.595 $\pm$ 0.029 & 0.092 $\pm$ 0.028 & 0.017 $\pm$ 0.020 & 26.3 $\pm$ 7.0 \\
10.4& 0.526 $\pm$ 0.042 & 0.107 $\pm$ 0.040 & 0.011 $\pm$ 0.024 & 15.7 $\pm$ 17.1 \\
    \hline
    \hline
    \end{tabular}
\caption{Ratios of helicity amplitudes $|T_{11}|/|T_{00}|$, 
  $|T_{01}|/|T_{00}|$, $|T_{10}|/|T_{00}|$ and phase difference 
  $\varphi_{11}-\varphi_{00}$, for five \qsq\ values, 
  extracted from the H1 and ZEUS measurements of the 15 combinations of spin density
  matrix elements~\protect\cite{h1_rho,zeus_rho_hq_me}.}
\label{table:val}
\end{table}


\end{document}